\documentclass[12pt,prd,aps,superscriptaddress,preprintnumbers,amsmath,amssymb,nofootinbib]{revtex4-1}
\usepackage{amsmath,graphics,epsfig}
\usepackage{color,hyperref}
\usepackage{datetime}
\usepackage{booktabs}
\begin{document}

\title{O' R Inflation in $F$-term Supergravity}
\author{ Sibo Zheng}
\affiliation{Department of Physics, Chongqing University, Chongqing 401331, P. R. China}
\date{October,  2016}

\begin{abstract}
The supersymmetric realization of inflation in $F$-term supergravity is usually plagued by the well known ``$\eta$" problem.
In this paper, a broad class of small-field examples is realized by employing general O' Raifeartaigh  superpotentials,
where the moduli is identified as the massless inflaton.
For illustration we present the simplest example in detail, which can be considered as a generalization of hybrid inflation.
\end{abstract}

\maketitle

\section{Introduction}

Inflation is an extensively studied and widely accepted scenario to solve both the flatness and horizon problems in the early Universe.
For a review, see, e.g. \cite{0907.5424}.
On the other hand, Supersymmetry (SUSY) is a scenario to deal with the divergent problem of Standard  Model Higgs mass,
which is inferred to probably play an important role in the high energy region (far above the weak scale) from both the null results 
at particle colliders (e.g., LHC) as well as dark matter direct detection experiments (e.g., LUX).
Therefore,  it is natural to ask the question whether inflation and SUSY are connected.

When we explore high energy physics near the Planck mass scale $m_{P}$,
gravitational effect has to be taken into account.
This implies that one should discuss the realization of such kind of inflation in the context of Supergravity.
Given the structure of potential $V_{\text{local}}=V^{F}_{\text{local}}+V^{D}_{\text{local}}$ in supergravity,
studies on the SUSY realizations of inflation can be divided into two classes -
the F-term and D-term supergravity.
In the former one, in unit of Planck mass
\begin{eqnarray}{\label{V}}
V^{F}_{\text{local}}=e^{K}\left[D_{i}W K^{-1}_{ij^{*}}D_{j^{*}}W^{*}-3\mid W\mid^{2}\right]
\end{eqnarray}
where $K(\Phi_{i})$ and $W(\Phi_{i})$ is the Kahler and superpotential, respectively,
and covariant derivative $D_{i}W=\partial_{i}W+K_{i}W$, 
with index $i$ referring to a chiral superfield $\Phi_{i}$.
The expotential factor $e^{K}$ introduces the well known ``$\eta$" problem to $F$-term Supergravity.
For reviews, see, e.g, \cite{9807278, 1101.2488}.
Progresses along this line are significantly improved by Kallosh {\it et al} \cite{1011.5945},
where the authors gave a prescription to construct a general potential of inflation,
but with the price of some specific shift symmetry as required in the Kahler potential.

There is no such ``$\eta$" problem in $D$-term supergravity \cite{9405389}.
Along this line,  it was firstly proposed by Halyo \cite{9606423} that 
a positive and non-zero $D$-term naturally realizes inflation.
Unfortunately, by following the insights in \cite{0904.1159}, 
the Fayet-Iliopoulos term itself is not consistent with the supergravity.

In this paper, we focus on the realization of inflation in $F$-term supergravity.
Unlike in \cite{1011.5945}, we turn to the canonical Kahler potential 
\footnote{$\eta$ problem which may be induced by operators with mass dimension higher than four in the Kahler potential will not be discussed here.} without any specific shift symmetry.
In order to keep our calculations perturbatively valid, 
we restrict to the small-field inflation\footnote{ For examples of the large-field inflation such as chaotic inflation, see, e.g., \cite{0004243, 1008.3375}.}.
Previous studies under these same assumptions in the literature include hybrid inflation \cite{9703209}, new inflation \cite{9608359} 
and single-field inflation \cite{1605.01934}.

The plan of this paper is organized as follows.
In Sec.II we give a prescription to construct a general superpotential without the $\eta$ problem for canonical Kahler potential.
We find that it is actually valid for  O' Raifeartaigh (O' R) superpotential.
In Sec.III we present a new example with the simplest O' R superpotential for illustration.
This model obviously differs from the hybrid inflation in the sense that there is no need of a critical mass scale $\sigma_{c}$ to end inflation
although it gives rise to a spectral index similar to the hybrid inflation.
Finally, we conclude in Sec.IV.

\section{O' R superpotential without  $\eta$ problem}
This section is devoted to a prescription, 
from which there is no $\eta$ problem for a renormalizable superpotential $W$ and canonical Kahler potential $K$.
In order to provide inflation the potential $V$ in Eq.(\ref{V}) must be positive and non-zero.
Consider that for the small-field inflation the local potential is well approximated to
\begin{eqnarray}{\label{limit}}
V^{F}_{\text{local}}\approx V^{F}_{\text{global}}+\mathcal{O}(1/m^{2}_{P}).
\end{eqnarray}
This requirement is equivalent to that the global SUSY should be broken $\mid F_{i}\mid^{2}\neq 0$.
Meanwhile, the inflaton mass should be always small \cite{9401011,1307.4343} (in compared with the Hubble parameter $H$)
 in order to evade the $\eta$ problem,
and the others scalars' masses should be comparable with $H$ if one considers a realization of single-field inflation 
\footnote{Note that $H^{2}\approx V_{\text{global}}/3$ in unit of Planck mass.}.
This requirement is equivalent to that there exists a flat direction with the modulus identified as the inflaton superfield $X$. 

In the global version of SUSY the two requirements above can be satisfied simultaneously by choosing general O' R superpotential,
\begin{eqnarray}{\label{superpotential}}
W(X, \chi_{i})= X(\mu^{2}+f(\chi_{i})) + g(\chi_{i}),
\end{eqnarray}
where $X$ and $\chi_i$ are chiral superfields,  and functions $f(\chi_{i})$ and $g(\chi_{i})$ are defined as
\footnote{In principle, linear terms like $a_{i}\chi_{i}$ and $b_{i}\chi_{i}$ may appear in $f(\chi_{i})$ and $g(\chi_{i})$, respectively. 
However, the former probably introduce dangerous mixing between $X$ and $\chi_{i}$, and the later introduce new linear structure except $X$ in the superpotential.
These terms can be forbidden by adjusting their $R$ charges.},
\begin{eqnarray}{\label{functions}}
f(\chi_{i}) &=& a_{ij}\chi_{i}\chi_{j},\nonumber\\
g(\chi_{i})&=& b_{ij}\chi_{i}\chi_{j}+b_{ijk}\chi_{i}\chi_{j}\chi_{k}.
\end{eqnarray}
The coefficients $a_{ij}$, $b_{ij}$, etc, are assumed to be real for simplicity,
and $\mu$ is the SUSY breaking scale.
The superpotential is only linear function of $X$,
from which with suitable $R$ charge assignments the minimal value of $V$ is determined by \cite{0703196},
\begin{eqnarray}{\label{vacuum}}
F_{X}\mid_{\left<\chi_{i}\right>} = \mu^{2},~~~~F_{\chi_{i}}\mid_{\left<\chi_{i}\right>}=0,~~~~ \text{and}~~~~X~ \text{arbitrary}
\end{eqnarray}

Now, let us examine whether there is indeed no $\eta$ problem in the local version in Eq.(\ref{V}) 
for superpotential in Eq.(\ref{superpotential}) with the structure of vacuum in Eq.(\ref{vacuum}).  
After a detailed calculation we find that \footnote{Keep in mind that operators with mass dimensions higher than 4 are compensated by the correct powers of Planck mass.},
\begin{eqnarray}{\label{Vnew}}
V^{F}_{\text{local}}=e^{K}\cdot \left\{A_{0}+[A_{1}X+h.c]+A_{2}\mid X\mid^{2}+[A_{3}X\mid X\mid^{2}+h.c]+A_{4}\mid X\mid^{4}\right\},
\end{eqnarray}
where
\begin{eqnarray}{\label{Ai}}
A_{0}&=&\mid F_{X}\mid^{2}+\mid F_{\chi_{i}}\mid^{2} + \mid g\mid^{2}(\mid \chi_{i}\mid^{2}-3)+[g^{*}F_{\chi_{i}}\chi_{i}+h.c]\nonumber\\
A_{1}&=&\left(F^{*}_{\chi_{i}}\chi^{*}_{i} +g^{*} \mid\chi_{i}\mid^{2}-2g^{*}\right)F_{X}\nonumber\\
A_{2}&=&-\mid F_{X}\mid^{2} + \mid F_{X}\mid^{2}\mid\chi_{i}\mid^{2}+\mid g\mid^{2}\nonumber\\
A_{3}&=&g^{*}F_{X}\nonumber\\
A_{4}&=& \mid F_{X}\mid^{4}.
\end{eqnarray}
Eq.(\ref{Vnew}) is organized in powers of $X$, 
the benefit of which is that the terms of mass squared for $X$ can be easily extracted from individual terms in Eq.(\ref{Vnew}) 
after one expands the expotential factor $e^{K}\approx 1+\mid X\mid^{2}+\mid\chi_{i}\mid^{2}$.

Under the small-field approximation ($\mid X\mid <<1$  etc.)
we obtain the final expression for $V^{F}_{\text{local}}$,
\begin{eqnarray}{\label{Vfinal}}
V^{F}_{\text{local}}&\approx& \tilde{A}_{0}+[\tilde{A}_{1}X+h.c]+\tilde{A}_{2}\mid X\mid^{2}+[\tilde{A}_{3}X\mid X\mid^{2}+h.c]+\tilde{A}_{4}\mid X\mid^{4}\nonumber\\
&+& [\tilde{A}_{5} X\mid X\mid^{4}+h.c] +\tilde{A}_{6} \mid X\mid^{6} +\cdots
\end{eqnarray}
where we have ignored higher-dimensional operators suppressed by $m_{P}$. 
Here, functions $\tilde{A}_{i}$ are given by,
\begin{eqnarray}{\label{tildeAi}}
\tilde{A}_{0}&=&(1+\mid\chi_{i}\mid^{2})A_{0},~~~~~~\tilde{A}_{1}=(1+\mid\chi_{i}\mid^{2})A_{1}  \nonumber\\
\tilde{A}_{2}&=&A_{0}+(1+\mid\chi_{i}\mid^{2})A_{2},~~\tilde{A}_{3}=A_{1}+(1+\mid\chi_{i}\mid^{2})A_{3} \nonumber\\
\tilde{A}_{4}&=&A_{2}+(1+\mid\chi_{i}\mid^{2})A_{4},~~\tilde{A}_{5}=A_{3},~~\tilde{A}_{6}=A_{4}
\end{eqnarray}
From Eq.(\ref{Vfinal}) the mass squared $m^{2}_{X}$ is read as,
\begin{eqnarray}{\label{Xmass}}
m^{2}_{X}=\left[\mid F_{X}\mid^{2}+\cdots\right]+(1+\cdots)\left[-\mid F_{X}\mid^{2}+\cdots\right]+\mathcal{O}(\mid X\mid/m_{P})
\end{eqnarray}
where we have used $F_{X}\mid_{\left<\chi_{i}\right>}=\mu^{2}$ and $\cdots$ represent contributions from high-dimensional operators.
Clearly, $m^{2}_{X}=0$ at the leading order,
and its smallness in compared with $H^{2}$ still holds as long as we restrict to the small-field inflation.

One may also verify that all the masses squared for $\chi_{i}$ are of order $H^{2}$ from $\tilde{A}_{0}$ in Eq.(\ref{tildeAi}) and Eq.(\ref{Ai}). 
It results from the fact that there is only one non-zero $F$-term in the minimal of $V$.

In summary, for small-field approximation our statements about masses for $X$ and $\chi_{i}$ are always true
as long as the mass scales of coefficients $a_{ij}$, $b_{ij}$, etc, in Eq.(\ref{functions}) are all of the order $\mu$ for  O' R superpotential.
In the light of our results it is easy to understand 
why there is no $\eta$ problem in some simple inflation models  such as hybrid inflation ($g=0$, $f=\bar{\chi}\chi$) in the literature.
We refer this broad class of small-field inflation models as O' R inflation.

\section{An Example}
By following the results in the previous section, 
in this section we propose a concrete and simple example for illustration.
As we will see, this new inflation model obviously differs from the hybrid inflation in the sense that there is no tachyon mass problem,
although it gives rises to a similar spectral index.

\subsection{Inflaton effective potential}
We begin with the simplest O' R model \cite{OR} in the class of superpotentials defined in Eq.(\ref{superpotential}),
\begin{eqnarray}{\label{OR}}
W=X(\mu^{2}+\frac{1}{2}h\chi^{2}_{1})+m\chi_{1}\chi_{2}.
\end{eqnarray}
In this model the SUSY-breaking vacuum is described by $\chi_{1,2}=0$ and arbitrary $X$,
and $V^{F}_{\text{global}}=\mid F_{X}\mid^{2}=\mu^{4}$.
After substituting $\left<f\right>\mid_{\chi_{1,2}=0}=0$ and $\left<g\right>\mid_{\chi_{1,2}=0}=0$ into Eq.(\ref{Vfinal})
one can verify that $m_{X}=0$.
So, $X$ is the candidate of inflaton.
Note that the structure of superpotential in Eq.(\ref{OR}) allows a 
$Z_2$ symmetry. In order to avoid possible production of domain wall \cite{DW} due to this $Z_2$ symmetry, 
we simply assume that it has been explicitly broken by higher dimensional operators \cite{0006168}.
It is expected that the domain wall problem can be resolved in more general O' R model without the aid of higher dimensional operators.

In order to derive the effective potential for $X$ we firstly calculate the mass spectral for the chiral superfields $\chi_{1,2}$.
By virtue of the standard formulas, the scalar masses squared $\mathcal{M}^{2}_{B}$ 
and the fermion masses squared $\mathcal{M}^{2}_{F}$ are given by, respectively,
\begin{eqnarray}{\label{scalarmatrix}}
\mathcal{M}^{2}_{B}= m^{2} \cdot
\left(%
\begin{array}{cccc}
  \mid\epsilon_{X}\mid^{2}+1 & \epsilon_{X} & \epsilon_{\mu} & 0 \\
 \epsilon^{*}_{X} &   1  & 0 & 0 \\
 \epsilon_{\mu} & 0 &  \mid\epsilon_{X}\mid^{2}+1 &  \epsilon^{*}_{X} \\
 0 & 0 & \epsilon_{X} & 1 \\
\end{array}%
\right)
\end{eqnarray}
and 
\begin{eqnarray}{\label{fermionmatrix}}
\mathcal{M}^{2}_{F}=m^{2} \cdot
\left(%
\begin{array}{cccc}
  \mid\epsilon_{X}\mid^{2}+1 & \epsilon_{X} & 0 & 0 \\
 \epsilon^{*}_{X} &   1  & 0 & 0 \\
0 & 0 &  \mid\epsilon_{X}\mid^{2}+1 &  \epsilon^{*}_{X} \\
 0 & 0 & \epsilon_{X} & 1 \\
\end{array}%
\right),
\end{eqnarray}
where $\epsilon_{X}=h\mid X\mid/m$ and $\epsilon_{\mu}=h\mu^{2}/m^{2}$.

Here a few comments are in oder regarding the magnitude of $\epsilon_{X}$ and $\epsilon_{\mu}$.
First, if $\epsilon_{\mu}<1$ fields $\chi_{i}$ would be decoupled from field $X$, inflation will nerve end. 
In the following analysis we will impose  $\epsilon_{\mu}\geq 1$.
Second, for $\mu$ is far below the Planck mass 
the ratio $\epsilon_{X}/\epsilon^{1/2}_{\mu}=h^{1/2}\mid X\mid/\mu$ is larger than unity,
which implies that the magnitude of $\epsilon_{X}$ is larger than unity as well.
It can be verified that for such $\epsilon_{X}$ and $\epsilon_{\mu}$ 
the signs of determinant of $\mathcal{M}^{2}_{B}$ in Eq.(\ref{scalarmatrix}) 
and that of $\mathcal{M}^{2}_{F}$ in Eq.(\ref{fermionmatrix}) are both positive,
and the effective potential for the inflaton ($\sigma=\mid X\mid$) can be approximated as,
\begin{eqnarray}{\label{effective}}
V_{\text{eff}}(\sigma)=\mu^{4}\left[1+\frac{h^{2}}{16\pi^{2}}\log\left(\frac{h\sigma}{\Lambda}\right) +\sigma^{6}+\cdots\right],
\end{eqnarray}
where $\Lambda$ denotes the renormalizable scale in the model,
and non-renormalizable terms with mass dimensions higher than $\sigma^{6}$ are neglected.

\subsection{Fit to inflationary parameters}
Now we discuss the implications in the simplest O' R inflation model.
First, we count the total number of e-folds during inflation,
\begin{eqnarray}{\label{N}}
N_{tot}=\int^{\sigma_{i}}_{\sigma_{end}} \frac{V}{V'} d\sigma= \int^{\sigma_{i}}_{\sigma_{d}} \frac{d\sigma}{6\sigma^{5}}  +\int^{\sigma_{d}}_{\sigma_{end}} \frac{16\pi^{2}}{h^{2}} \sigma d\sigma 
\simeq \pi^{4/3}h^{-4/3}\left[\left(\frac{1}{144}\right)^{1/3}+\left(\frac{16}{3}\right)^{1/3}\right],\nonumber\\
\end{eqnarray}
where $\sigma_{i}$ and $\sigma_{end}$ is the initial and end value of inflaton (in unit of Planck mass), respectively,
and $\sigma_{d}\simeq (h^{2}/96\pi^{2})^{1/6}$ being the critical value 
above (below) which the non-renormalizable term (log-term) dominates the inflation.
Eq.(\ref{N}) shows that the period between $\sigma_{in}$ and $\sigma_{d}$ gives rise to  $\sim10\%$ contribution to $N_{tot}$.
From Eq.(\ref{N}) the requirement of the e-fold number $N_{tot}=50(60)$ then leads to $h=0.27(0.24)$.

Second, consider the period between $\sigma_{d}$ and $\sigma_{end}$.
The slow roll parameters are given by,
\begin{eqnarray}{\label{slowrole}}
\epsilon &=& \frac{1}{2} \left(\frac{V'}{V}\right)^{2}\simeq \frac{h^{2}}{64\pi^{2}N},\nonumber\\
\eta &=& \frac{V''}{V} \simeq -\frac{1}{2N},
\end{eqnarray}
where $N$ is the e-fold number corresponding to the inflaton value $\sigma_{N}=\sqrt{N h^{2}/8\pi^{2}}$ during inflation.
From Eq.(\ref{slowrole}) one finds that 
$i)$ the magnitude of $\mid\epsilon\mid$ is small in comparison with $\mid \eta\mid$ due to the one-loop factor suppression.
$ii)$, Unlike the hybrid inflation where $\sigma_{end}$ is approximately determined by the critical value $\sigma_{c}$, 
for our model where there is no critical mass scale corresponding to tachyon mass 
$\sigma_{end}$ is determined to be $h/4\pi$ by taking $\mid\eta_{end}\mid=1$.
To summarize, $\sigma_{d}\simeq 0.21$ and $\sigma_{end}\simeq 0.02$ for $N_{tot}=50$.

\begin{figure}
\includegraphics[width=0.5\textwidth]{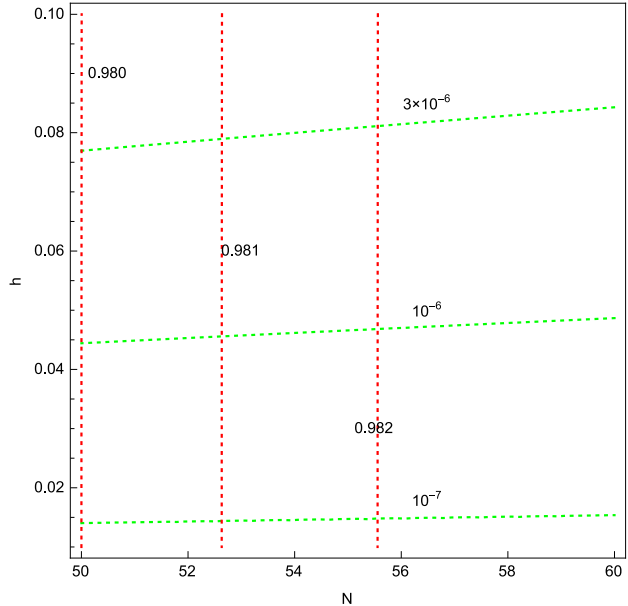}
\vspace{-0.5cm}
 \caption{Parameter space in the plane of $h-N$. the curves of $r$ and $n_{s}$ are shown in green and red, respectively.
 Only Regions between the right-hand side of the red line $n_{s}=0.98$ and left-hand side of the red line $n_{s}=0.982$ are allowed.}
\label{nr}
\end{figure}

In terms of the slow roll parameters in Eq.(\ref{slowrole}) the main inflationary parameters measured by experiments are given by,
\begin{eqnarray}{\label{index}}
n_{s}&=&1-6\epsilon+2\eta \simeq 1-\frac{1}{N}, \nonumber\\
\frac{dn_{s}}{d\text{ln k}} &\simeq& 16\epsilon\eta-24\epsilon^{2}\simeq -\frac{h^{2}}{8\pi^{2}N^{2}}, \nonumber\\
r&=&16\epsilon \simeq \frac{h^{2}}{4\pi^{2}N},\nonumber\\
\mathcal{R}_{s}&=& \frac{1}{24\pi^{2}} \frac{V}{\epsilon} \simeq \frac{8\mu^{4}N}{3h^{2}},
\end{eqnarray}
where $n_{s}$ is the spectral index, $dn_{s}/d\text{ln k}$ is the running, $r$ is the tensor-to-scalar ratio,
and $\mathcal{R}_{s}$ is the amplitude of primordial fluctuations.
In Fig.\ref{nr} we show the parameter space by fitting to the experimental constraints on the inflationary parameters in Table.\ref{data}.
It shows that only regions between the right-hand side of the solid red line $n_{s}=0.98$ and left-hand side of the dotted red line $n_{s}=0.982$ are allowed.
We would like to mention that the spectral index in this model is mainly sensitive to the total number of e-folds during inflation.
Typically, we have $n_{s}\sim 0.98$, $r\sim 10^{-6}$ and $dn_{s}/ d\text{ln k} \sim 10^{-7}$.

\begin{table}
\begin{center}
\begin{tabular}{cc} 
  & $\Lambda \text{CDM+r+nrun}$
\\
\hline\hline
 $n_{s}$ &  $ 0.9721\pm 0.011$ (68\% CL) \\
  $\frac{dn_{s}}{d lnk} $ & $-0.0038 \pm 0.0068$ (68\% CL)  \\
  $r_{0.01}$  & $\leq 0.075$ (95\% CL)  \\
  $\text{ln}(10^{10}\mathcal{R}_{s})$  & $3.117 \pm 0.021$ (68\% CL)  \\
  \hline
\end{tabular}
\caption{The latest experimental limits at 68\% CL on the inflationary parameters in the cosmological model $\Lambda \text{CDM+r+nrun}$ 
from $P15+BK14+BAO15$ data combination \cite{1512.07769}.}
\label{data}
\end{center}
\end{table}

Let us examine the constraints on magnitude of $h$.
Substituting the experimental value $\mathcal{R}\simeq 2.44\times 10^{-9}$ in Table \ref{data} into the last formula in Eq.(\ref{index}) gives rise to $\mu \simeq 2.0\times 10^{-3}h^{1/2}$ for $N=50$,
which together with $\epsilon_{X}\geq 0.1$ and $\epsilon_{\mu}\geq 1$ implies $\sigma_{end}\geq 2.5\times 10^{-4}$.
At the meantime, $h$ is fixed to $0.24-0.27$ as required by $N_{tot}=50-60$,
which is consistent with this bound.

Finally, we would like to mention that there remain a few issues to be addressed for a complete phenomenological study.
First, it is likely to tune the value of $n_{s}$ from the expectation value of hybrid inflation ($\sim 0.98$) to 
the favored value of Planck data ($\sim 0.974$ \cite{Planck}) by taking the effect of reheating into account. 
See, e.g., \cite{1606.06677}.
Second, the tuning of initial condition and gravitino overproduction problem usually plague SUSY driven inflation model. 
See Refs. \cite{1007.5152,1404.1832} for similar discussions in hybrid inflation.
These aspects are beyond the scope of this paper.

\section{Conclusion}
A prescription to construct a general superpotential without the $\eta$ problem for canonical Kahler potential has been addressed in the $F$-term supergravity.
We have verified that it is valid for O' R superpotential at least for the small-field inflation.
Also, the simplest O' R model is studied in detail for illustration,
which is found to be similar to the hybrid inflation.
Since there is no need of a critical mass scale $\sigma_{c}$ to end the inflation,
it can be considered as a generalization of hybrid inflation.
Although it is not a concrete statement, 
the constraint on the spectral index in the simplest O' R model may be relaxed in some non-minimal examples.

\begin{acknowledgments}
The author would like to thank the referee for comments.
This work is supported in part by the Fundamental Research Funds for the Central Universities with project No. 106112016CDJXY300004.
\end{acknowledgments}

\end{document}